# Electrical control for extending the Ramsey spin coherence time of ion-implanted nitrogen-vacancy centers in diamond

S. Kobayashi[1,3], Y. Matsuzaki[2], H. Morishita[3], S. Miwa[1,4,5], Y. Suzuki[1,4], M. Fujiwara[3], N. Mizuochi[3,4,*]

[1]*Graduate School of Engineering Science, Osaka University, Toyonaka, Osaka 560-8531, Japan*

[2] *Device Technology Research Institute, National institute of Advanced Industrial Science and Technology (AIST), Central2, 1-1-1 Umezono, Tsukuba, Ibaraki 305-8568, Japan*

[3]*Institute for Chemical Research, Kyoto University, Uji, Kyoto 610-0011, Japan*

[4]*Center for Spintronics Research Network (CSRN), Osaka University, Toyonaka, Osaka 560-8531, Japan*

[5]*The Institute for Solid State Physics, The University of Tokyo, Kashiwa, Chiba 277-8581, Japan*

**Abstract**

The extension of the spin coherence times is a crucial issue for quantum information and quantum sensing. In solid state systems, suppressing noises with various techniques have been demonstrated. On the other hand, an electrical control for suppression is important toward individual controls of on-chip quantum information devices. Here we show the electrical control for extension of the spin coherence times of 40 nm-deep ion-implanted single nitrogen vacancy center spins in diamond by suppressing magnetic noises. We applied 120 V DC across two contacts spaced by 10 micrometers. The spin coherence times, estimated from a free-induction-decay and a Hahn-echo decay, were increased up to about 10 times (reaching 10 microseconds) and 1.4 times (reaching 150 microseconds), respectively. From the quantitative analysis, the dominant decoherence source depending on the applied static electric field was elucidated. The electrical control for extension can deliver a sensitivity enhancement to the DC sensing of temperature, pressure and electric (but not magnetic) fields, opening a new technique in solid-state quantum information devices.

[*]Corresponding author: mizuochi@scl.kyoto-u.ac.jp

I. INTRODUCTION

In quantum sensing [1-21] and quantum information science [22-26], long coherence times are crucially important. The improvement in sensitivity is inversely proportional to the square root of the coherence times [1-9]. Additionally, their increase directly improves quantum memory times [1-5,22-25] and quantum gate fidelity [26,27] in quantum information devices. Among solid state systems, a negatively charged nitrogen-vacancy (NV) center in diamond is significantly interested because it has long coherence times [1-5,22] and high sensitivities with a nanometer scale resolution [6-15]. It has been eagerly investigated to realize novel applications such as high-precision magnetic [6,7,16] and electric field sensing [8-11,17], thermometry [12,13], sensing in/on living cell [13-15], quantum metrology [18,19], pressure sensing [20,21], and quantum information processing [22-25].

Among the techniques to produce the NV centers for quantum sensing, an ion-implantation technique is the most effective way to deliberately put NV centers very close to the surface with nanoscale resolution [28], which is needed for nanoscale sensing of matters outside the diamond as well as for having the best spatial resolution for sensing [7]. However, it is well known that the ion-implanted NV centers have much shorter coherence times compared with the best ones due to defects created during the ion implantation [7, 29, 30]. Therefore, technique to improve the spin coherence time is important.

The extension of the spin coherence time has been demonstrated by suppressing noises with various techniques such as tailored dynamical decoupling [31-35], measurements at low temperature [3], measurements at high magnetic field [36], and decoupling by fast charge-state changes with high power laser irradiation [22] in addition to removal of noise sources by growth techniques [1,2,4,5]. On the other hand, realization of the suppression by an electric field is important [37-39] because the electric field can be locally controlled in individual on-chip devices such as scalable quantum device in a dense array [37-39] and quantum sensing device. Furthermore, it does not need huge energy consumption facilities for operation and rare materials synthesized by isotopes without a nuclear spin.

The electrical control is challenging because the electric fields do not couple directly to the spin unlike the magnetic field. Previously, the static electric field dependence on magnetic resonance frequencies of the NV center was reported [17]. By using the dependence, application of the NV center to the nano-scale electric field sensors was demonstrated [8,10]. In the previous research [8], the analysis of the electronic structure of the NV center reveals how an applied magnetic field influences the electric-field-sensing properties. In this study, coherence times, which were estimated from a free-induction-decay ($T_2^{\text{FID}}$) and a Hahn-echo decay ($T_2^{\text{echo}}$), were measured in the externally applied static electric field. We report the increase of $T_2^{\text{FID}}$ and $T_2^{\text{echo}}$ under the electric field and discuss the mechanism.

II. EXPERIMENTAL METHOD

The NV center is a well-studied defect in diamond. It consists of a substitutional nitrogen atom adjacent to a vacancy site. The spin triplet of the ground state exhibits zero-field splitting of $D_{gs}/h \approx 2.87$ GHz between the $m_s = 0$ and degenerate $m_s = \pm 1$ spin sub-levels, where $h$ is the Planck constant. The spin state of the triplet can be polarized, and readout by an optical excitation and a photoluminescence (PL) intensity measurement. Owing to these properties, the electron spin resonance of the NV center can be sensed by optically detected magnetic resonance (ODMR) techniques.

In our experiment, NV centers were created by ion implantation and subsequent annealing on a IIa (100) single-crystalline diamond substrate. Nitrogen-14 ($^{14}$N) with a natural abundance concentration (99.6 %) was implanted at a depth of about 40 nm by an ion-implantation with a kinetic energy of 30 keV. Electrodes and an antenna to apply an electric field and microwaves, respectively, were formed on the substrate (Figure 1(a)). From the current–voltage property of the structure to apply an electric field (Figure 1(b)), it was supposed that the voltage at the interface between the diamond and the electrode was zero and the electric field strength in diamond bulk was linearly proportional to the applied electric field (see Appendix A). The direction and strength of the electric field in diamond were simulated numerically (Figure 1(c)). The dots marked by 'NV1', 'NV2', and 'NV3' indicate the positions of the NV centers. The estimated direction of the electric field for each NV center is almost parallel to the y-axis with a polar angle $\theta_E$ of 89° or 91° and an azimuth angle $\varphi_E$ of 90° or 270° in the NV coordinate system (Figure 1(d)). As shown in Figure 1(e), the NV axes of NV1, NV2, and NV3 are parallel to one the following axis: $[\bar{1}\bar{1}1]$, $[111]$, $[11\bar{1}]$ or $[\bar{1}\bar{1}\bar{1}]$, respectively.

### III. RESULTS AND DISCUSSION

Figure 2(a) shows the magnetic resonance spectra of NV1 obtained by ODMR, while Figure 2(b) plots the resonance frequencies from the fit with six Gaussian functions. Since six resonance signals are observed at each electric field, it seems that the NV electron spin interacts with a nuclear spin of $^{14}$N (I = 1). Thus, the z-component of the external magnetic field $B_z$ can be estimated from the obtained resonance frequencies at the zero electric field. The estimated $B_z$ values for NV1, NV2, and NV3 are 13 μT, 12 μT, and 11 μT, respectively. The magnetic fields are likely from a residual field and/or Earth magnetism. As we increase the electric field, the spectral width becomes narrower and the resonance frequency shift becomes larger. From the narrowing of the spectral width, an increase in the coherence time by the electric field is inferred. Regarding the shifts in the resonance frequencies by the electric field, fittings based on the spin Hamiltonian of the ground triplet state were used in the quantitative analysis. The Hamiltonian is expressed as

$$H_{gs} = \frac{1}{\hbar^2}\left(D_{gs} + d_{gs}^{\parallel} E_z\right)\left[S_z^2 - \frac{1}{3}S(S+1)\right] + \frac{1}{\hbar^2}d_{gs}^{\perp}\left[E_x\left(S_y^2 - S_x^2\right) + E_y\left(S_x S_y + S_y S_x\right)\right] + \frac{1}{\hbar}\mu_B g_e \mathbf{S} \cdot \mathbf{B}$$

(1)

where $\hbar$, $\mu_B$, and $g_e$ are the reduced Planck constant, the Bohr magneton, and the g factor of the electron spin, respectively. **S** is a spin operator and the spin quantum number of the NV center is 1 (S=1). $d_{gs}^{\parallel}/h =$ $0.35 \pm 0.02$ kHz·cm/kV and $d_{gs}^\perp/h = 17 \pm 3$ kHz·cm/kV [17] are the measured axial and non-axial components of the ground triplet state electric dipole moment, respectively. To consider the effect of the $^{14}$N nuclear spin, the Hamiltonian describing the hyperfine interaction with the nuclear spin and the quadrupole interaction of the $^{14}$N is introduced to the spin Hamiltonian (1). The introduced Hamiltonian is given as

$$H_{\text{hfc}} = \frac{1}{\hbar^2}\left[A_\parallel S_z I_z + A_\perp (S_x I_x + S_y I_y) + P I_z^2\right]$$

(2)

where $A_\parallel/h = -2.1$ MHz and $A_\perp/h = -2.7$ MHz [40] are the axial component and the non-axial component of the hyperfine interaction, respectively. $I$ represents the nuclear spin of $^{14}$N, and $P/h = -5.0$ MHz [39] is the strength of the quadrupole interaction of the $^{14}$N nuclear.

The resonance frequencies at each electric field were determined by a numerical calculation and fitted to experimental results, where the free parameter of the fitting is the non-axial components of the electric dipole moment $d_{gs}^\perp$. $B_z$ was set to the estimated value from the spectrum at the zero electric field. $B_x$ and $B_y$ were set to zero because their contributions are negligible under the conditions of $D_{gs} \gg \mu_B g_e B$ and $D_{gs} \gg d_{gs}^\perp E_\perp$.

Figure 2(b) shows the fitted curves as a solid line. The curves fit well with the experimental results, and $d_{gs}^\perp$ is estimated to be $d_{gs}^\perp/h = 19$ kHz·cm/kV, 16 kHz·cm/kV, and 16 kHz·cm/kV for NV1, NV2, and NV3, respectively. The estimated values agree with the measured value of $d_{gs}^\perp/h = 17 \pm 3$ kHz·cm/kV within the margin of error. These results indicate that the spin Hamiltonian given by equation (1) is valid in the range of a high electric field to ~100 kV/cm for the non-axial component of the electric field effect on the ground state electron spin of the NV center. Additionally, we measured the shifts of the resonance frequencies at a range of ± ~100 kV/cm. The shifts behave as even functions with respect to the applied electric field. The result, indicating no electric field polarity dependence of the shifts, is consistent with the spin Hamiltonian (see Appendix B).

It is worth mentioning that, in Figures 2 (a) and 2(b), the hyperfine splitting observed in the ODMR seems to be weaker as we increase the amplitude of the applied electric fields. This effect can be understood as follows. The hyperfine coupling effect on the NV center from the nuclear spin can be interpreted as effective Zeeman energy induced by magnetic fields from the nuclear spin. However, under the electric fields, the Zeeman splitting due to the magnetic fields will be suppressed. This means that the hyperfine splitting is also suppressed by applying the strong electric fields.

In the previous research [41], it is shown that the frequency shifts of the magnetic resonance become dependent on the transverse orientation of the electric/magnetic field (i.e. $\phi_E$ and $\phi_B$, where $\tan\phi_E = E_y/E_x$, and $\tan\phi_B = B_y/B_x$) if the electric and the magnetic fields perpendicular to the N-V axis ($E_\perp$ and $B_\perp$) are simultaneously applied. They derived the equation for this dependence in the case that $E_\perp$ and $B_\perp$ are simultaneously applied, which is valid in the limit $\mu_B^2 B_\perp^2/2D_{gs} \gg d_{gs}^\perp E_\perp$.

In our research, the geomagnetic field is unintentionally applied although we do not know the direction of the geomagnetic field with respect to the N-V axis. Even if its direction is in the transverse plane, the applied electric field in our experiment is strong enough to satisfy the relation $\mu_B^2 B_\perp^2 / 2D_{gs} \ll d_{gs}^\perp E_\perp$. As shown in their paper, in the presence of either an electric filed or a magnetic field, the resonance frequencies depend on the alignment of the electric/magnetic field with the center's major symmetry axis (i.e. $\theta_E$ and $\theta_B$), but do not depend on the transverse orientation of the electric/magnetic field (i.e. $\phi_E$ and $\phi_B$), where $\tan\theta_E = E_\perp / E_z$ and $\tan\theta_B = B_\perp / B_z$. Therefore, in our experiment, we can assume that the resonance frequencies does not depend on $\phi_E$ and $\phi_B$.

To measure the coherence time, the microwave frequency was set to the arrowed transition shown in Figure 2(a), and applying a weak microwave selectively excited the resonance. The resonance is related to the $^{14}$N nuclear spin state of $m_I = 0$, where the effect of the hyperfine interaction on the resonance frequency is quite small. In this study, we characterized two kinds of coherence times $T_2^{FID}$ and $T_2^{echo}$. To measure the FID and the Hahn echo signal, the Ramsey sequence and the Hahn echo sequence, depicted in Figures 3(a) and 4(a), respectively, were employed. Figures 3(b) and 4(b) show the decay curves of the PL intensity as a function of $\tau$, corresponding to the loss of coherence obtained from NV1. Both $T_2^{FID}$ and $T_2^{echo}$ increase as the external electric field strength increases. $T_2^{FID}$ and $T_2^{echo}$ increase up to ~10 times and 1.4 times, respectively. The decay curves of the free induction decay of NV2 and NV3 are shown in Appendix C.

The dynamics of the NV electron spin in bulk diamond are influenced by the locally fluctuating magnetic field induced by the surrounding spins (the spin bath) [2,42]. The impact of the spin bath on an NV electron spin is limited to dephasing due to a fluctuating magnetic field along the NV's quantization axis [31,36]. The magnetic field noise is represented by adding an effective noise term of $\mu_B g_e b_z(t)$ where $b_z(t)$ denotes the magnetic field from the spin bath. We assume that the spin bath can have temporal fluctuations, and so we consider $b_z(t)$ as random magnetic fields that stochastically changes in time. We modeled the magnetic field fluctuation by the Ornstein-Uhlenbeck (O-U) process with the auto-correlation function, $C^b(t) = \langle b_z(0) b_z(t) \rangle = b_z^{\sigma 2} \exp(-|t|/\tau_c^b)$, where $\tau_c^b$ is the correlation time and $b_z^{\sigma 2}$ is the standard deviation of the amplitude distribution of the magnetic field fluctuation. The details of the spin bath are described in Appendix D. In the regime of $\sqrt{(\mu_B g_e B_z)^2 + (d_{gs}^\perp E_\perp)^2} \gg \mu_B g_e b_z^\sigma$, which corresponds to our experiments that have a strong bias field or strong transverse electric field, a fluctuation in the resonance frequency $\delta\omega(t)$ induced by the magnetic field fluctuation under an electric field is expressed as

$$\delta\omega(t) = \sqrt{\left(\mu_B g_e (B_z + b_z(t))\right)^2 + \left(d_{gs}^\perp E_\perp\right)^2} - \sqrt{(\mu_B g_e B_z)^2 + \left(d_{gs}^\perp E_\perp\right)^2} \cong R^b \cdot \frac{\mu_B g_e b_z(t)}{\hbar}$$

(3)

where $R^b(E_\perp) = \frac{\mu_B g_e B_z}{\sqrt{(\mu_B g_e B_z)^2 + (d_{gs}^\perp E_\perp)^2}}$. $T_2^{FID}$ and $T_2^{echo}$ are estimated by the fit with functions of $f^{FID}(\tau) = y_0 + A \cdot \exp(-(\tau/T_2^{FID})^2)$ for FID and $f^{echo}(2\tau) = y_0 + A \cdot \exp(-(2\tau/T_2^{echo})^3)$ for the Hahn echo, which are characteristics of a slowly fluctuating spin bath with $\frac{\hbar}{R^b \mu_B g_e b_z^\sigma} \ll \tau_c^b$ (see Appendix E). From these equations, we obtained the curves by fitting the data. Although there is a gap between the data and the curves because of the scatter of the data, the obtained fitted curves are represented as blue solid lines in Figures 3(b) and 4(b), while the estimated $T_2^{FID}$ and $T_2^{echo}$ are plotted in Figures 3(c) and 4(c), respectively. Note that in Figures 3(b) and 4(b), the electric field strength is normalized by the magnetic field along the z-axis (i.e., the horizontal axis represents $d_{gs}^\perp E_\perp/\mu_B g_e B_z$). Both $T_2^{FID}$ and $T_2^{echo}$ increase in accordance with the increase in the applied electric field.

To confirm the mechanism for the increase of the coherence time, we fitted the functions

$$T_{2,b_z}^{FID}(E_\perp) = \frac{\sqrt{2}\hbar}{R^b \mu_B g_e b_z^\sigma}$$

(4)

$$T_{2,b_z}^{echo}(E_\perp) = \left(12\tau_c^b \left(\frac{\hbar}{R^b \mu_B g_e b_z^\sigma}\right)^2\right)^{\frac{1}{3}}$$

(5)

to the experimental results of $T_2^{FID}$ and $T_2^{echo}$ with respect to the electric field, where the free parameter is $b_z^\sigma$ for the fit of $T_2^{FID}$ and $\tau_c^b$ for the fit of $T_2^{echo}$. $b_z^\sigma$ in the fit of $T_2^{echo}$ was set to the value estimated in the fit of $T_2^{FID}$. Note that we supposed $b_z^\sigma$ and $\tau_c^b$ are independent of the applied electric field. The fitted curves are represented as the dashed curves for NV1 and the chained curve for NV2 in Figures 3(c) and 4(c), respectively. Regarding $T_2^{FID}$, the fitted curves basically reproduce the experimental results. The estimated $b_z^\sigma$ values are 6 μT and 5 μT for NV1 and NV2, respectively.

In contrast to $T_2^{FID}$, the fitted curve for $T_2^{echo}$ does not reproduce the experimental results. Particularly in the high field region, the experimental result of $T_2^{echo}$ is almost saturated. However, the fitted curve does not show saturation. Instead, the fitted curve increases even at a high electric field. Therefore, another decoherence source should be considered in the analysis of $T_2^{echo}$.

For the additional decoherence source, the fluctuation in the electric field should be considered [43-46]. Although the influence of the fluctuation in the electric field on the resonance frequency becomes small in the regime of $\mu_B g_e B_z \gg d_{gs}^\perp E_\perp$ [45], the influence of the electric field fluctuation becomes larger than that of the magnetic field fluctuation in the regime of $\mu_B g_e B_z \ll d_{gs}^\perp E_\perp$. This behavior is consistent with the experimental result of $T_2^{echo}$. Therefore, the fluctuation in the electric field is a candidate for the additional decoherence source. In the case of our system, the relevant fluctuations come from the y-component of the electric fields due to the following reasons. First, the z-component of the fluctuation can be ignored because the axial component of the electric dipole moment $d_{gs}^\parallel$ is ~50 times smaller than the non-axial component $d_{gs}^\perp$. The x-component can be ignored because the fluctuation in the non-axial electric field $E_\perp = \sqrt{{E_x}^2 + {E_y}^2}$ induced by the *x*-component of the electric field fluctuation is suppressed in our experimental condition ($E_x \ll E_y$).

We modeled the electric field fluctuation $e_y(t)$ by the O-U process with the auto-correlation function, $C^e(t) = \langle e_y(0)e_y(t)\rangle = {e_y^\sigma}^2 \exp(-|t|/\tau_c^e)$, where $\tau_c^e$ is the correlation time and ${e_y^\sigma}^2$ is the standard deviation of the magnitude of the fluctuation. In addition, we assumed that the fluctuations of the magnetic field and the electric field are not correlated. In the regime of $\sqrt{(\mu_B g_e B_z)^2 + (d_{gs}^\perp E_\perp)^2} \gg \mu_B g_e b_z^\sigma,\ d_{gs}^\perp e_y^\sigma$, the fluctuation in the resonance frequency $\delta\omega(t)$ due to the fluctuations of the magnetic and electric fields can be represented as

$$\delta\omega(t) = \sqrt{\left(\mu_B g_e(B_z + b_z(t))\right)^2 + \left(d_{gs}^\perp(E_\perp + e_y(t))\right)^2} - \sqrt{(\mu_B g_e B_z)^2 + (d_{gs}^\perp E_\perp)^2}$$

$$\cong R^b \cdot \frac{\mu_B g_e b_z(t)}{\hbar} + R^e \cdot \frac{d_{gs}^\perp e_y(t)}{\hbar}$$

(6)

where $R^e(E_\perp) = \frac{d_{gs}^\perp E_\perp}{\sqrt{(\mu_B g_e B_z)^2 + (d_{gs}^\perp E_\perp)^2}}$. Given the slow electric field fluctuation with $\frac{\hbar}{R^e d_{gs}^\perp e_y(t)} \ll \tau_c^e$, the electric field dependences of $T_2^{\text{FID}}$ and $T_2^{\text{echo}}$ are derived as

$$T_2^{FID}(E_\perp) = \frac{\sqrt{2}\hbar}{\sqrt{(R^b \mu_B g_e b_z^\sigma)^2 + (R^e d_{gs}^\perp e_y^\sigma)^2}}$$

(7)

$$T_2^{echo}(E_\perp) = \frac{T_{2,b_z}^{echo} \cdot T_{2,e_y}^{echo}}{\left(T_{2,b_z}^{echo\,3} + T_{2,e_y}^{echo\,3}\right)^{\frac{1}{3}}}$$

(8)

where $T_{2,e_y}^{echo} = \left(12\tau_c^e \left(\frac{\hbar}{R^e d_{gs}^\perp e_y^\sigma}\right)^2\right)^{\frac{1}{3}}$. Regarding $T_2^{\text{FID}}$, it seems that the influence of the fluctuation in the electric field is negligible compared to that of the magnetic field fluctuation, even at $E_\perp \sim 100$ kV/cm, because the increase of $T_2^{\text{FID}}$ by the electric field is not saturated. Thus, the fluctuation in the electric field cannot be estimated by fitting with function (7) as the fitting is meaningless. However, we can roughly estimate the upper limit of the amplitude of the electric field fluctuation $e_y^\sigma$. We assumed that the upper limit is the value satisfying the influence of the electric field fluctuation on $T_2^{\text{FID}}$, which is ten times smaller than the influence of the magnetic field fluctuation [i.e., $(R^e d_{gs}^\perp e_y^\sigma)^2 \lesssim \frac{1}{10}(R^b \mu_B g_e b_z^\sigma)^2$]. The estimated value is $e_y^\sigma \lesssim 0.5$ kV/cm at $E_\perp \sim 100$ kV/cm.

Regarding $T_2^{\text{echo}}$, we fitted function (8) to the experimental results with respect to the strength of the electric field, where the free parameters are $\tau_c^b$ and $\tau_c^e/e_y^{\sigma 2}$. Figure 4(c) shows the fitted curve as a solid line, reproducing the experimental results well. The estimated free parameters are $\tau_c^b$ = 170 ms and $\tau_c^e/e_y^{\sigma 2} = 6\ \mathrm{ms\cdot cm^2/kV^2}$. From the estimated value and the upper limit of $e_y^\sigma$, the upper limit of $\tau_c^e$ is estimated to be ~1.4 ms. The estimated values of the correlation times, $\tau_c^b$ = 170 ms and $\tau_c^e$ < 1.4 ms, are more than ten times larger than the time scale of $T_2^{\mathrm{echo}}$ ~ 0.1 ms, which does not conflict with the assumptions of the slow fluctuations of the magnetic field and the electric field used in this analysis.

We discuss the source of electric field fluctuation. The reported surface charge fluctuation [43, 44, 46] is a potential source of the electric field fluctuation. Considering a point charge $q$ on the surface, the resulting electric field at the NV center in the diamond is expressed as $\mathbf{E} = \frac{1}{4\pi\varepsilon_0}\frac{2}{\kappa_d+\kappa_{oil}}\frac{q}{r^2}\hat{\mathbf{r}}$ [47], where the screening effect is not counted. $r$ is the distance between the surface electron and the NV center, and $\hat{\mathbf{r}}$ is unit vector in the direction of the NV center. $\kappa_d = 5.7$ and $\kappa_{oid} = 2.3$ are the dielectric constants of diamond and optical immersion oil, respectively. $\varepsilon_0$ is the permittivity of free space. For a single elementary charge located immediately above (40 nm) the NV center, the strength of the electric field is ~2 kV/cm. Although the value is larger than the upper limit of $e_y^\sigma$ estimated above, this result is not necessarily incongruous because the electric field is partially screened by the interplay of the electrons in the diamond lattice, maintaining electroneutrality. We consider that the origin of the electric field fluctuation may be the surface charge fluctuation [44] or charge fluctuation of impurities/defects in bulk. Our results indicate that $T_2^{\mathrm{echo}}$ will increase almost linearly as shown as dotted line in Fig. 4 (c) by reducing the fluctuation.

In our case, the maximum applied electric field is limited by a breakdown of the electrode deposited on diamond. This upper limit will be improved technically by finding optimized fabrication processes and will be extended to the intrinsic break-down voltage of diamond. As far as we know, one of the largest break down voltage of diamond is reported to be more than 2 MV/cm [48]. It is more than 10 times larger than the maximum applied electric field (166 kV/cm) in the present experiment. Therefore, in principle, a further extension of $T_2^*$ can be expected.

**IV. CONCLUSION**

The electric field dependence of $T_2^{\mathrm{FID}}$ and $T_2^{\mathrm{echo}}$ at a range up to ~100 kV/cm were quantitatively analyzed based on the spin Hamiltonian. $T_2^{\mathrm{FID}}$ and $T_2^{\mathrm{echo}}$ increase up to ~10 times and 1.4 times, respectively. Assuming only a magnetic field fluctuation, the behavior of $T_2^{\mathrm{FID}}$ is basically elucidated. The behavior of $T_2^{\mathrm{echo}}$ is well reproduced assuming fluctuations in both the electric and magnetic fields. Although the magnetic field fluctuation is the dominant decoherence source for $T_2^{\mathrm{FID}}$ in the entire range of the electric field in our experiment and for $T_2^{\mathrm{echo}}$ in the low electric field region, the dominant decoherence source for $T_2^{\mathrm{echo}}$ in the high electric field region is the electric field fluctuation. The difference in the dominant decoherence source under the electric field is due to the difference in the amplitude–frequency characteristics, corresponding to the difference in the correlation time of the magnetic field fluctuation and the electric field fluctuation in our experiment. The enhancement of the coherence times by the electric field can contribute to the improvement of the sensitivity in thermometry, pressure and AC electric field sensing, although it is not effective to the magnetic field sensitivity [44]. The present technique can be utilized for not only for the NV center in diamond but also the high-spin centers (S>1) with the zero-field splitting, such as promising centers in silicon carbide [49-51]. Our study opens up the new technique of the electrical decoupling of the spin coherence in solid from the magnetic noises.

**Appendix A.  Materials and application of the electric field**

The diamond substrate was a CVD-grown IIa (100) single-crystalline diamond purchased from Element Six Corporation (electronic grade). The nitrogen impurity concentration was less than 5 ppb and the concentration of $^{13}$C was the natural abundance in the diamond. $^{14}$N with a natural abundance concentration was ion-implanted at the Ion Technology Center Co., Ltd with a kinetic energy of 30 keV, implanting nitrogen at the depth of about 40 nm with the density about $5\times10^8$/cm$^2$. Subsequent annealing at 800 °C for 1 hour using rapid thermal annealing equipment produced the NV centers with the density around less than 0.1 ppb. To remove the residual surface contamination and terminate the surface with oxygen, the sample was kept in a mixture of H$_2$SO$_4$ and HNO$_3$ at 200 °C for 1 hour. Ti(30 nm)/Au(100 nm) electrodes for the electric field and an antenna for the microwaves were formed on the substrate by conventional electron beam lithography and metal deposition processes.

The linear current–voltage (I–V) property of the structure for applying electric field was obtained although there were Schottky barriers at the interface between the electrodes and diamond. Schottky barriers exhibit linear I–V property in the regime of $V \ll \left|\frac{kT}{e}\right|$, where $k$ is the Boltzmann constant, $T$ is the temperature, and $e$ is the elementary charge. Thus, in the range of the applied electric field, the voltage at the interfaces appeared to be much smaller than $\left|\frac{kT}{e}\right|$ (~0.026 V at RT) The voltage at the interface was negligible compared to the applied voltage to the electrodes. Therefore, we assumed that the voltage at the interface was zero and the electric field strength in diamond bulk was linearly proportional to the applied electric field.

The direction and strength of the electric field in diamond was simulated numerically (Figure 1(c)). The dots marked by 'NV1', 'NV2', and 'NV3' indicate the positions of the NV centers of NV1, NV2, and NV3, respectively. Their positions were estimated from the confocal scan images and the depth of implanted nitrogen. The direction of the N–V axes of the NV centers was parallel to the one with $[111], [\bar{1}\bar{1}\bar{1}], [11\bar{1}],$ or $[\bar{1}\bar{1}1]$, which was confirmed by the measurements of the PL intensity with polarized light. From the results, the direction of the electric field in the NV coordinate system depicted in Figure 1(d) was estimated to be almost parallel to the $y$-axis with the polar angle $\theta_E$ of 89° or 91° and the azimuth angle $\varphi_E$ of 90° or 270°.

**Appendix B.  Electric field polarity dependence of the resonance frequency shifts and the transition rates**

We measured the electric field polarity dependence of the resonance frequency shifts at a range of ± ~ 100 kV/cm because asymmetry of the optical transition depending on the polarity was reported previously [52]. Measured NV center was another one, NV4. The direction of the N–V axis of NV4 was parallel to the one with $[1\bar{1}1]$, or $[\bar{1}1\bar{1}]$, which is different from NV1-3. Thus, the *x*-component of the electric field is much larger than the *y*-component. Figure 5(a) shows the magnetic resonance spectra of NV4 obtained by ODMR. While four resonance signals are observed at the zero electric field, the two dips at the middle of the four dips are larger than the others. It seems that the two resonance signals overlap at the middle two dips and the NV electron spin interacts with a nuclear spin of $^{14}$N (I = 1). Then the *z*-component of the external magnetic field $B_z$ can be estimated from the obtained resonance frequencies at the zero electric field. The estimated $B_z$ value is 34 µT. Figure 5(b) plots the resonance frequencies of the middle two with respect to the electric field strength. The shifts behaved as even functions with respect to the applied electric field. The result, indicating no electric field polarity dependence of the shifts, is consistent with the spin Hamiltonian.

Regarding the size of the dips, an electric field polarity dependence is observed. This may be due to a change of the magnetic transition rates. The transition rates are approximately described by Fermi's golden rule as $\frac{2\pi}{\hbar}|\langle S_\pm|H_{MW}|S_0\rangle|^2$, where $|S_0\rangle$, $|S_+\rangle$, and $|S_-\rangle$ are eigenstates of the electron spin which are expressed as $|S_0\rangle = |0\rangle$, $|S_+\rangle = e^{i\frac{\varphi_E}{2}}\cos\frac{\theta}{2}|+1\rangle - e^{-i\frac{\varphi_E}{2}}\sin\frac{\theta}{2}|-1\rangle$, and $|S_-\rangle = e^{i\frac{\varphi_E}{2}}\sin\frac{\theta}{2}|+1\rangle + e^{-i\frac{\varphi_E}{2}}\cos\frac{\theta}{2}|-1\rangle$, respectively. Here $\tan\theta = \frac{d^\perp_{gs}E_\perp}{\mu_B g_e B_z}$ and $|0\rangle$, $|+1\rangle$, and $|-1\rangle$ are eigenstates of $S_z$. $H_{MW}$ is the perturbing Hamiltonian due to the microwave. In our measurement system, the *x*-component of the oscillating field from the microwave $B_{MW\_x}$ is much larger than the *y*-component. Thus, the perturbing Hamiltonian is expressed as $H_{MW} = \frac{1}{\hbar}\mu_B g_e S_x B_{MW\_x}$. In the case where the electric field is along the *y*-axis, the transition rates don't depend on the strength of the electric field. Actually, the observed Rabi frequencies of the NV1 at each electric field were almost same. On the other hand, in the case where the electric field is along the *y*-axis, the transition rates depend on the strength of the electric field. This result agrees with the experimental results of NV4 qualitatively.

**Appendix C.  Decay curves of the free induction decay of NV2 and NV3**

The decay curves of the free induction decay of NV2 and NV3 are shown in Figures 6(a) and 6(b), respectively. Ramsey pulse sequence to measure $T_2^{FID}$ is the same with that of NV1 as shown in Figure 3(a). Fit function of $f^{FID}(\tau) = y_0 + A \cdot \exp(-(\tau/T_2^{FID})^2)$ is employed. The estimated $T_2^{FID}$ are plotted in Figure 3(c).

**Appendix D.   Consideration of the spin bath**

The main decoherence source of the NV centers located in bulk diamond is the fluctuation in the magnetic field induced by paramagnetic centers in the lattice [2,42]. For the NV centers situated near the surface (<10 nm), the main sources of the decoherence are the fluctuations in the magnetic field due to the flip of electron-spins at the surface [29] and the fluctuations in the electric field due to a charge fluctuation in the surface states[43,44]. In our experiments, the NV centers are placed around 40 nm from the surface. Therefore, the fluctuation in the magnetic field from paramagnetic centers should be the main source of decoherence. With regard to the paramagnetic centers, the nuclear spin of $^{13}$C atoms and the electron spins of some kind of defects (e.g. N-substituted defects) are possible candidates. For a case in which naturally abundant $^{13}$C (1.1 %) is a main source of decoherence, $T_2^{echo}$ for the NV electron is about 500 μs [2].

In our experiment, the time scale of the decay curves of the Hahn echo measurement is ~ 100 μs. The short time scale in our experiments indicates that the addition contribution to the decoherence is attributed to the electron spins, constituting a spin bath. Since the magnetic moments of the electron spins in the spin bath and NV centers are similar in size, there is no significant difference in the coupling strength of between intra-bath spin–spin interaction and bath-spin to the NV electron spin interaction. In such a situation where many spins interacting each other with similar coupling strength, the effect of the NV electron spin to the spin bath diffuses to the bath rapidly. This behavior is similar to the individual spins in the bath, allowing the fluctuation in the magnetic field to be treated as stochastic noise made by a classical Markov process [31,36]. Additionally, under the assumption of a uniform distribution of the defects in the spin bath, the fluctuation in the magnetic field is expressed by a Gauss distribution. In summary, the magnetic field fluctuation according to the stationary Ornstein-Uhlenbeck (O-U) process is assumed in this study.

Although the fluctuation in the magnetic field is three-dimensional, only the *z*-component affects the fluctuation in the resonant frequency and decoherence. The *x*- and *y*-components flip the NV electron spin, in principle. However, a large difference in the spin flipping energy between NV center and other defects due to the zero-field splitting of 2.87 GHz in the NV center results in negligible spin-flip probabilities [31,36]. Here, we denote the *z*-component of the fluctuating magnetic field as $b_z(t)$. From the aforementioned assumptions, the distribution of $b_z(t)$ is a Gaussian with an average value of $\langle b_z(t) \rangle = 0$. The auto-correlation function $C^b(t)$ is $C^b(t) = \langle b_z(0) b_z(t) \rangle = b_z^{\sigma 2} \exp(-|t|/\tau_c^b)$. Here, $\tau_c^b$ is the correlation time and $b_z^{\sigma 2}$ is the standard deviation in a fluctuating magnetic field. It should be noted that the correlation time $\tau_c^b$ and standard deviation $b_z^{\sigma 2}$ correspond to the strength of the spin–spin interaction inside the spin bath and the NV center to the bath-spin interaction, respectively.

**Appendix E.   The analysis of $T_2^{FID}$ and $T_2^{echo}$**

To derive the fitted functions for electric field dependence of $T_2^{FID}$ and $T_2^{echo}$, the time evolutions of the spin in the Ramsey and Hahn echo sequences with the fluctuation in the resonance frequency should be calculated. The statistical average with respect to the fluctuations should be taken into account. The population of the NV electron spin with $m_\mathrm{s} = 0$ after free evolution was expressed as $\langle p_{S_0}^{FID}(\tau)\rangle = \frac{1}{2} - \frac{1}{2}\mathrm{Re}\left[\langle \exp(-i\int_0^\tau \delta\omega(t)dt)\rangle\right]$ for the FID and $\langle p_{S_0}^{echo}(2\tau)\rangle = \frac{1}{2} + \frac{1}{2}\mathrm{Re}\left[\langle \exp(-i\int_0^\tau \delta\omega(t)dt)\exp(i\int_\tau^{2\tau}\delta\omega(t)dt)\rangle\right]$ for the Hahn echo. In the case where the fluctuation in the resonance frequency is expressed by (3), $\langle p_{S_0}^{FID}(\tau)\rangle$ and $\langle p_{S_0}^{echo}(2\tau)\rangle$ were derived as $\langle p_{S_0}^{FID}(\tau)\rangle = \frac{1}{2} - \frac{1}{2}\exp\left[-\tau_c^{b\,2}\left(R^b \cdot \frac{\mu_B g_e b_z^\sigma}{\hbar}\right)^2\left(\frac{\tau}{\tau_c^b} - 1 + e^{-\frac{\tau}{\tau_c^b}}\right)\right]$ and $\langle p_{S_0}^{echo}(2\tau)\rangle = \frac{1}{2} + \frac{1}{2}\exp\left[-\tau_c^{b\,2}\left(R^b \cdot \frac{\mu_B g_e b_z^\sigma}{\hbar}\right)^2\left(\frac{2\tau}{\tau_c^b} - 3 - e^{-\frac{2\tau}{\tau_c^b}} + 4e^{-\frac{\tau}{\tau_c^b}}\right)\right]$, respectively. Given the slow magnetic field fluctuation with $\frac{\hbar}{R^b \mu_B g_e b_z^\sigma} \ll \tau_c^b$, $\langle p_{S_0}^{FID}(\tau)\rangle$ and $\langle p_{S_0}^{echo}(2\tau)\rangle$ were simplified to be $p_{S_0}^{FID}(\tau) = \frac{1}{2} - \frac{1}{2}\exp\left(-\frac{1}{2}\left(R^b \cdot \frac{\mu_B g_e b_z^\sigma}{\hbar}\right)^2 \tau^2\right)$ and $p_{S_0}^{echo}(2\tau) = \frac{1}{2} + \frac{1}{2}\exp\left(-\frac{1}{12\tau_c^b}\left(R^b \cdot \frac{\mu_B g_e b_z^\sigma}{\hbar}\right)^2 (2\tau)^3\right)$. From these equations, the functions describing the electric field dependence of $T_2^{FID}$ and $T_2^{echo}$ were derived as equations (4) and (5). Equations (7) and (8) were derived by the same calculations.

**ACKNOWLEDGEMENTS**

The authors acknowledge financial support by KAKENHI (No. 15H05868, 16H02088, 16H06326) and partially by CREST (JPMJCR1333), MEXT Q-LEAP (No. JPMXS0118067395)). S. K. and N. M. acknowledge T. Shimo-Oka for helpful discussion.

**Additional information**

The authors declare no competing financial interests.

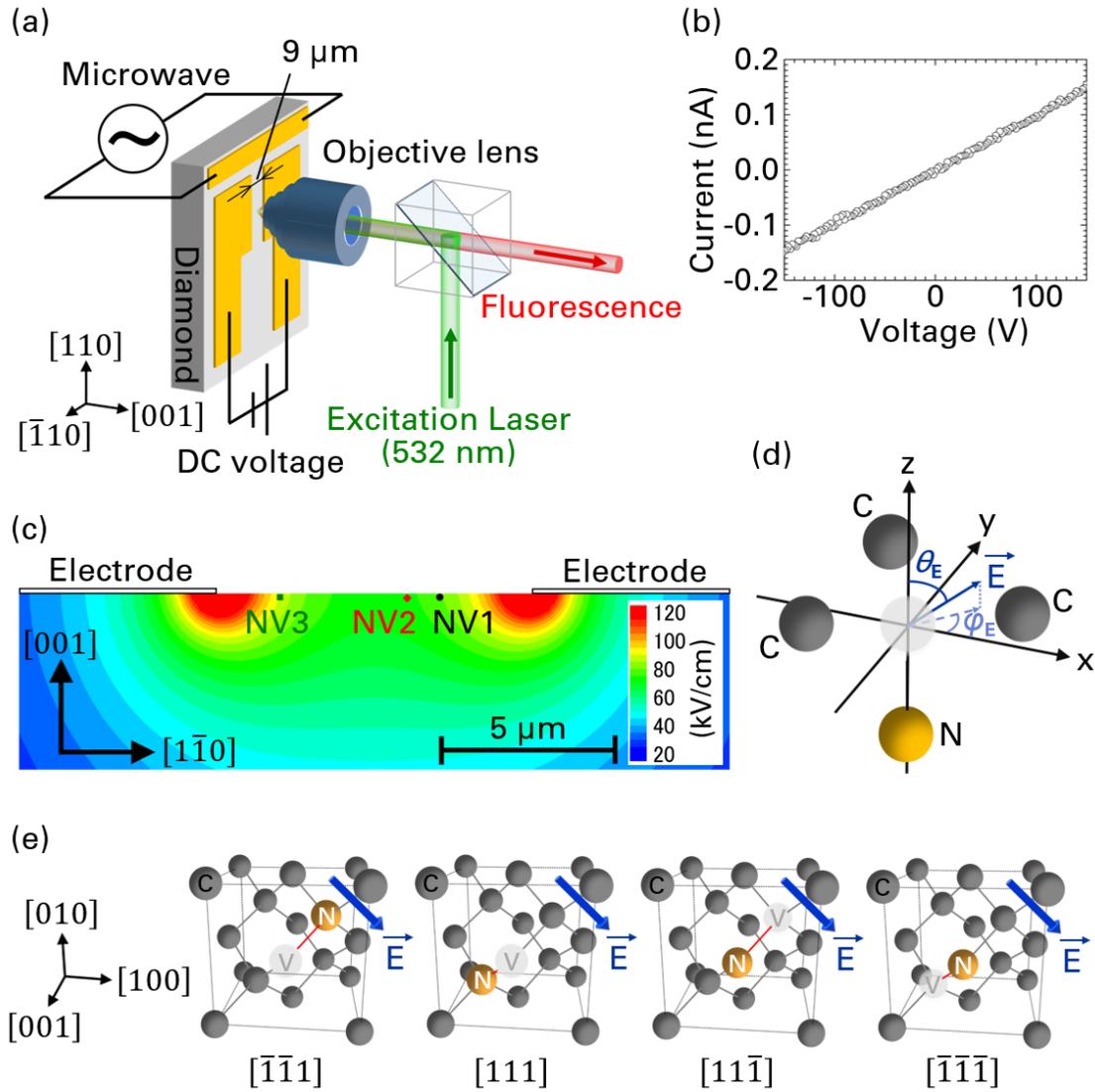

Figure 1. (a) Schematic image of the measurement setup. (b) Current-voltage property between the electrodes on the sample. Linear I-V property is observed. (c) Distribution of the electric field strength in the case where 100 V between the electrodes. Dots marked as NV1, NV2, and NV3 indicate the positions of the NV centers of NV1, NV2, and NV3, respectively. (d) Coordinate system of the NV centers. (e) Schematic images of the directions of NV axes. The NV axes of NV1, NV2, and NV3 are parallel to one of the following axis: [1̄1̄1], [111], [11̄1̄] or [1̄1̄1̄], respectively.

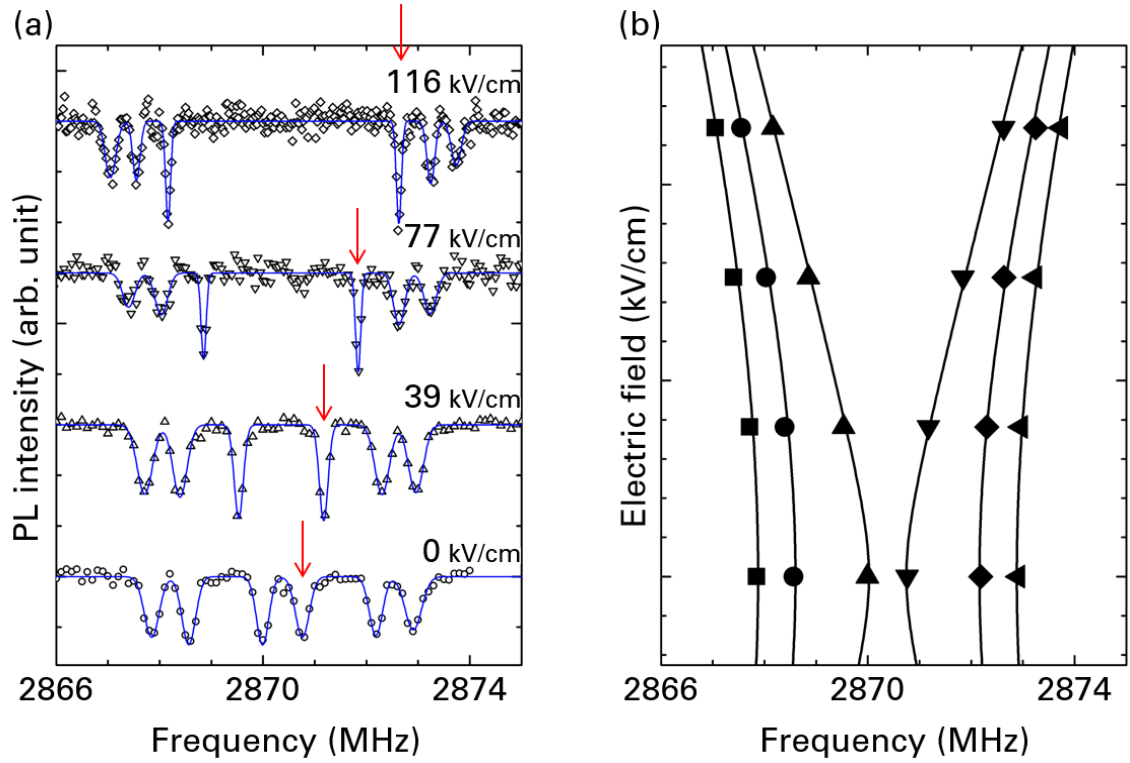

Figure 2. (a) ODMR spectra under various electric fields. Plots represent the experimental results of NV1. Solid lines represent the fitted curves using six Gaussian functions. (b) Electric field dependence of the resonance frequencies. Plots represent the estimated resonance frequencies from the experimental results of NV1. Solid lines represent the fitted curves calculated from the spin Hamiltonian of the ground state electron as (1) with the hyperfine interaction and quadrupole interaction of $^{14}$N nuclear as (2).

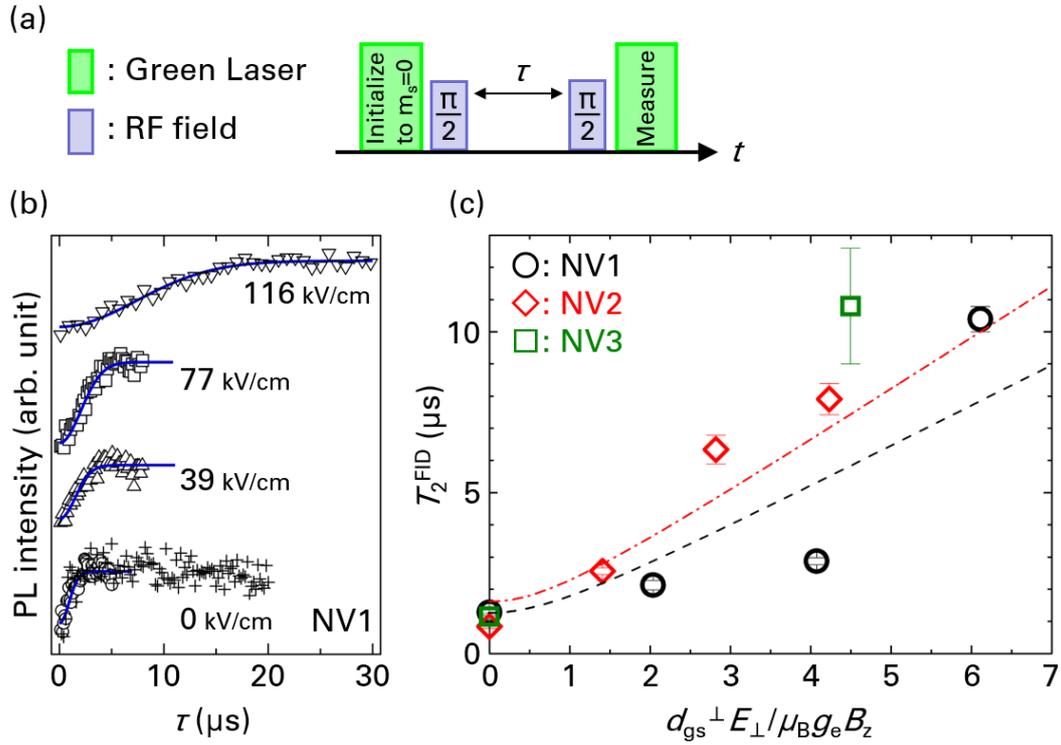

Figure 3. (a) Ramsey pulse sequence to measure $T_2^{\mathrm{FID}}$. (b) Decay curves of the free induction decay of NV1. Fit function of $f^{FID}(\tau) = y_0 + A \cdot \exp(-(\tau/T_2^{FID})^2)$ is employed. In 0 kV/cm, two data are represented as circle and plus. (c) Electric field dependence of $T_2^{\mathrm{FID}}$. Horizontal axis, which represents the electric field strength, is normalized by the magnetic field along the $z$-axis. Dashed curve represents the fitted curve for NV1. Chained curve represents the fitted curve for NV2.

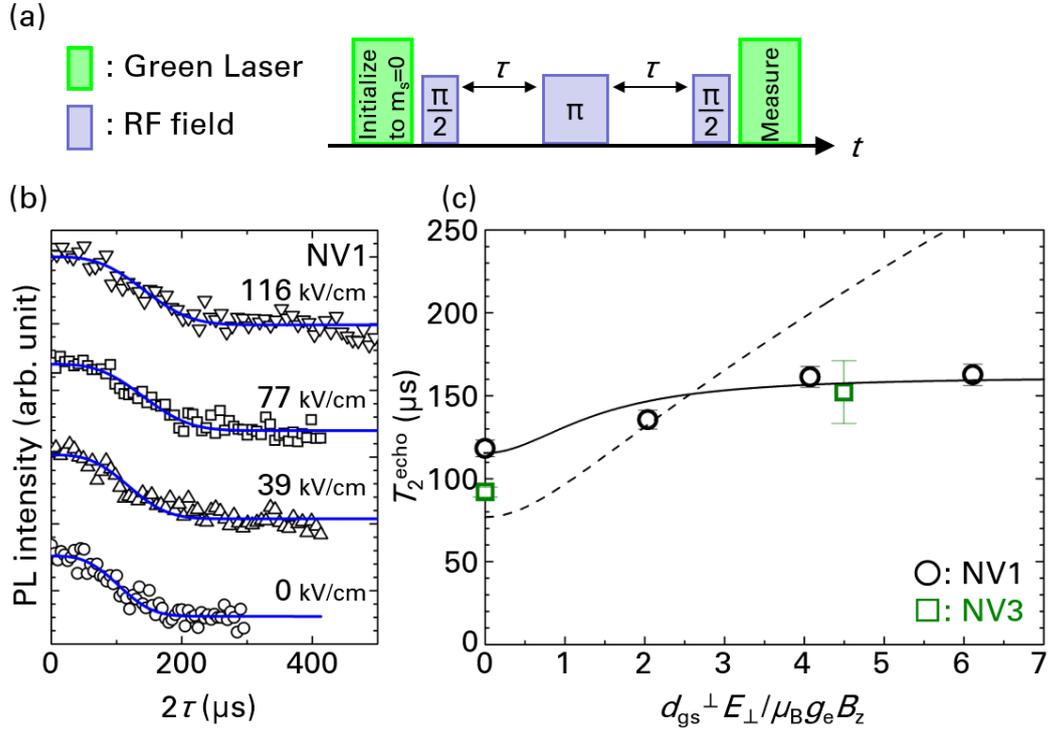

Figure 4. (a) Hahn echo pulse sequence to measure $T_2^{\text{echo}}$. (b) Decay curves of the Hahn echo signal of NV1. Fit function of $f^{echo}(2\tau) = y_0 + A \cdot \exp(-(2\tau/T_2^{echo})^3)$ is employed. (c) Electric field dependence of $T_2^{\text{echo}}$. Horizontal axis, which represents the electric field strength, is normalized by the magnetic field. Dashed curve represents the fit curve with magnetic field fluctuation only. Solid curve represents the fit curve with the magnetic and electric field fluctuations.

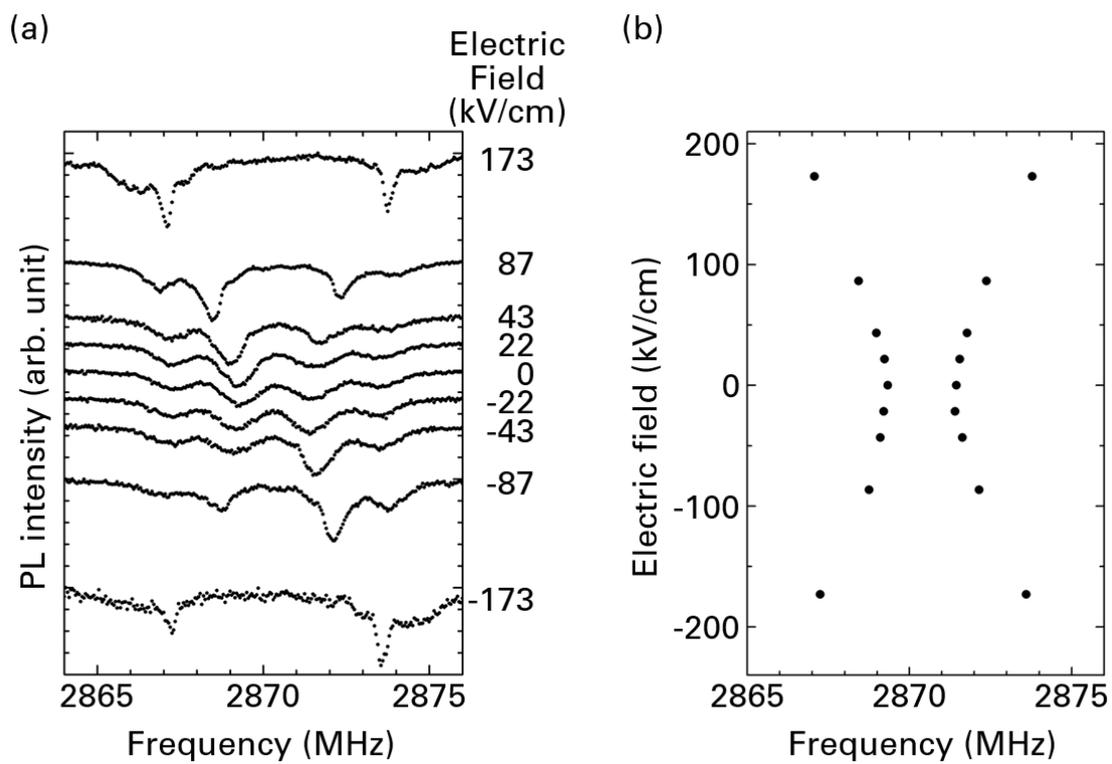

Figure 5. (a) ODMR spectra under various electric fields. Plots represent the experimental results of NV4. (b) Electric field dependence of the resonance frequencies. Plots represent the estimated resonance frequencies from the experimental results of NV4.

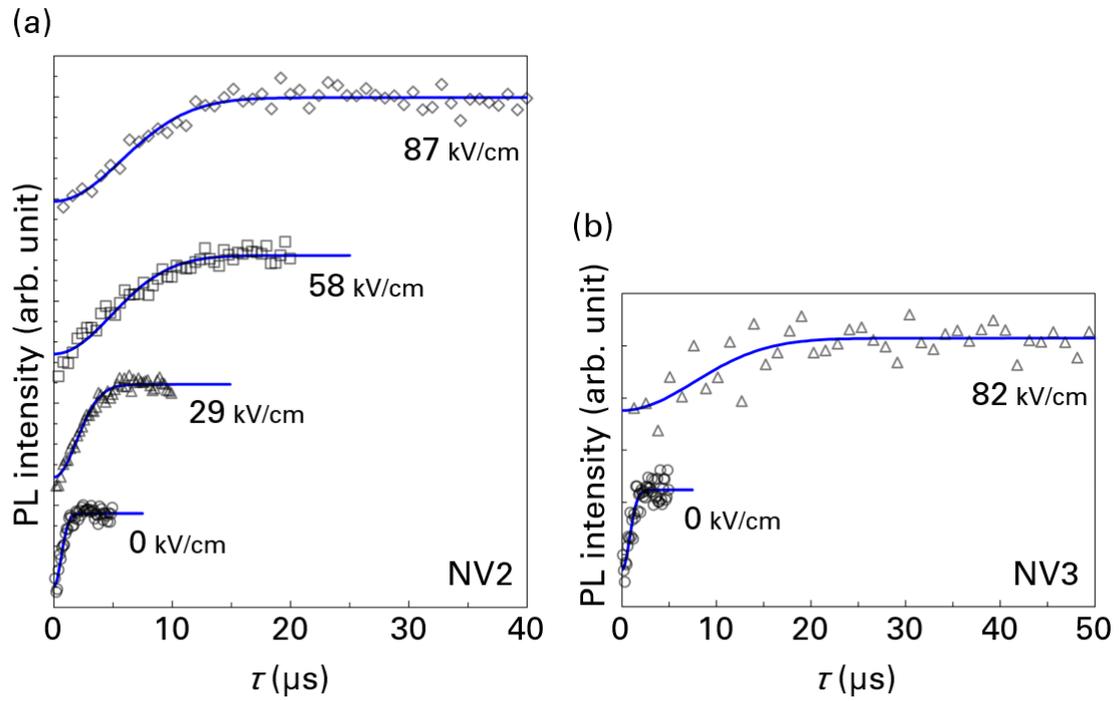

Figure 6. (a) Decay curves of the free induction decay of NV2. Fit function of $f^{FID}(\tau) = y_0 + A \cdot \exp(-(\tau/T_2^{FID})^2)$ is employed. (b) Decay curves of the free induction decay of NV3. Fit function of $f^{FID}(\tau) = y_0 + A \cdot \exp(-(\tau/T_2^{FID})^2)$ is employed.